\documentclass[aip,jap,twocolumn,reprint]{revtex4}
\usepackage{hyperref}
\usepackage{graphicx,epsfig,color}
\usepackage{longtable}
\long\def\symbolfootnote[#1]#2{\begingroup%
\def\thefootnote{\fnsymbol{footnote}}\footnote[#1]{#2}\endgroup}

\newcommand{\0}{\phantom{0}}
\newcommand{\1}{\phantom{$^\prime$}}
\newcommand{\2}{\phantom{-}}

\begin{document}

\title{Structural, electronic, vibrational and dielectric  properties of 
LaBGeO$_5$ from first principles}

\author{R. Shaltaf}
\email[E-mail me at: ]{r.shaltaf@ju.edu.jo}
\author{H. K. Juwhari}
\author{B. Hamad}
\author{J. Khalifeh}
\affiliation{
Department of Physics,
the University of Jordan,
11942  Amman, Jordan
}

\author{G.-M. Rignanese}
\author{X. Gonze}
\affiliation{
Institute of Condensed Matter and Nanosciences (IMCN/NAPS)\\
Universit\'e Catholique de Louvain (UCL), 8 Chemin des \'etoiles, B-1348 Louvain-la Neuve, Belgium.\\
European Theoretical Spectroscopy Facility (ETSF)
}

\date{\today}

\begin{abstract}

Structural, electronic, vibrational and dielectric properties of LaBGeO$_5$ with the stillwellite structure are determined based on \textit{ab initio} density functional theory.
The theoretically relaxed structure is found to agree well with the existing experimental data with a deviation of less than 0.2\%. 
Both the density of states and the electronic band structure are calculated, showing five distinct groups of valence bands.
Furthermore, the Born effective charge, the dielectric permittivity tensors, and the vibrational frequencies at the center of the Brillouin zone are all obtained.
Compared to existing model calculations, the vibrational frequencies are found in much better agreement with the published experimental infrared and Raman data, with absolute and relative rms values of 6.04~cm$^{-1}$, and 1.81\%, respectively.
Consequently, numerical values for both the parallel and perpendicular components of the permittivity tensor are established as 3.55 and 3.71 (10.34 and 12.28), respectively, for the high-(low-)frequency limit.

\end{abstract}

\maketitle

\section{INTRODUCTION}

Natural stillwellite~\cite{McAndrew,Juwhari} is a rare earth mineral, which accepts a very wide range of substitutions, described by the general formula (\textit{Ln}$^{+3}$,M$^{+2}$)B(Si,Ge,Al,P)(O,OH,F)$_{5.0}$ where \textit{Ln}=La, Ce, Pr, or Nd. 
Among these, LaBGeO$_5$ (LBG) can be taken as a prototype to represent the series of the larger rare-earth borogermanate stillwellite compounds.
LBG is well-known for its fascinating ferroelectric and nonlinear optical properties.~\cite{Vouagner1,Vouagner2}

LBG is characterized by its low ferroelectric transition temperature, Tc = 530$^\circ$C.
It exhibits a significant pyroelectric coefficient $\sim$5--10 nC cm$^{-2}$ K$^{-1}$, a low dielectric constant ($\sim$11),
a dielectric loss of ($\tan\delta$$\sim$0.001) at room temperature,
a second harmonic generation efficiency (SHG) of $\sim$30 units of
$\alpha$ quartz.
It also maintains a high electric resistance up to $\sim$500$^{\circ}$C.~\cite{Stefanovich,Onodera,Sigaev,Belokoneva-97}
Due to these appealing interesting properties, LBG has drawn considerable interest in recent years.
Some of its applications are a self-doubling laser~\cite{Capmany-1998,Capmany-1997} and more recently a substrate for growing high-quality crystalline InN thin films.~\cite{Miyazawa}

The LBG crystal has a trigonal stillwellite (CeBSiO$_5$)-type structure, 
with a $P_{31}$ space group symmetry and three formula units (Z=3) 
per unit cell. The principal structural units of stillwellite consist 
of infinite helical chains of BO$_4$ tetrahedra running parallel to the threefold screw axis, 
with each three tetrahedra forming a ring. 
Meanwhile, the GeO$_4$ tetrahedra were found to be connected to the remaining free vertices of one adjacent boron 
tetrahedron and to the lanthanide polyhedra.

Several groups have studied the polarized and unpolarized infrared (IR) and Raman spectra of LBG.~\cite{Rulmont,Kaminskii,Pisarev,Hruba}
Hrub\'a \textit{et al.},~\cite{Hruba} carried out an investigation 
of the Infrared (IR) and Raman inelastic scattering in the temperature range of 300 to 870 K. 
They reported the vibrational frequencies and the frequency-dependent dielectric function. 
On the other hand, based on a short-range potential functional calculation model, Smirnov \textit{et al.}~\cite{Smirnov} 
reported the IR and Raman spectral active modes of LBG. 

In the present study, we extend such previous results, and present detailed calculations of the 
vibrational and dielectric properties of LBG using the state of the art ab initio density functional
and density functional perturbation theories.
The agreement between the experimental reflectivity and our computed reflectivity, shown in Fig.3, is excellent.
To the best of our knowledge, there had been no previous first-principle study of LBG properties,
be they electronic or vibrational.

After the present introductory section, we detail our method of calculation (Sec. II), and we present our results (Sec. III), 
concerning respectively the structural properties, the electronic properties, the Born effective charge,
the vibrational and dielectric properties. We then conclude (Sec. IV).

\begin{figure*}[htb]
\begin{center}
\includegraphics{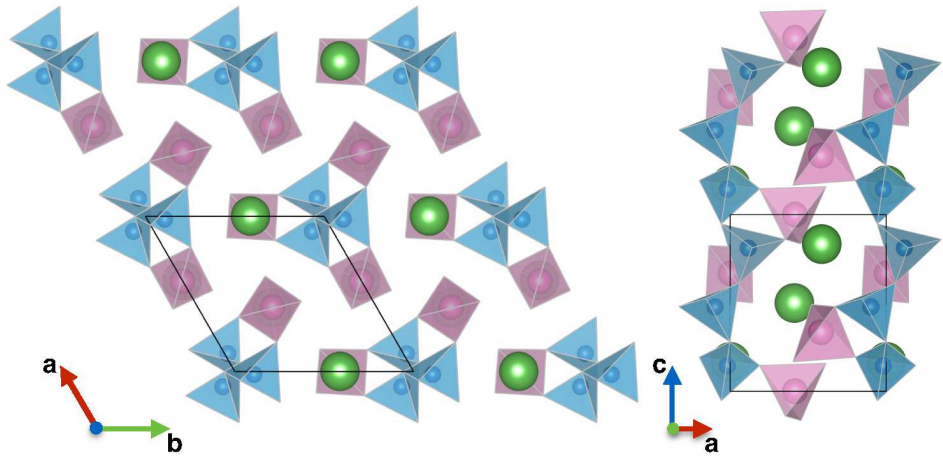}
\end{center}
\caption{Schematic representation of the LaBGeO$_5$ stillwellite crystalline structure: top (left panel) and side (right panel) views.
La atoms are in green, Ge tetrahedra in pink, and B tetrahedra in blue.
\label{figure1}}
\end{figure*}

\section{Method of Calculations}

The calculations have been performed using the state of the art \textit{ab initio} plane-wave density functional method, as implemented in the ABINIT code.~\cite{ ABINIT}
The all-electron potential is approximated by norm-conserving pseudopotentials generated within the Troullier-Martin\cite{TM} 
and extended Teter schemes.\cite{Teter}
The pseudopotentials are constructed considering the following states as valence
states: La($5s^2, 5p^6, 6s^2, 5d^1$), Ge($3s^2,3p^6$), B($3s^2,3p^1$) and O($2s^2,2p^4$).
The wave functions are expanded into a plane-wave basis set up to a kinetic
energy cutoff of 40 Ha.  The Brillouin zone integration is performed using a
special 4$\times$4$\times$4 grid of $k$-points generated with the Monkhorst-Pack
scheme.~\cite{ Monkhorst-Pack} 
We have checked that the computations are well converged with these numerical parameters.
The density of states calculations are carried out using the tetrahedron method 
using a 16$\times$16$\times$16 grid of $k$-points. 
The electron-electron interaction is approximated within local-density approximation (LDA).~\cite{PW92} 
Relaxations of the lattice parameters and internal atomic positions within
the unit cell were performed using the BFGS algorithm~\cite{BFGS} until the maximum component
of the force acting on any atoms has dropped below
$10^{-6}$~Hartree/Bohr and until the maximal stress is lower than $10^{-5}$~Hartree/Bohr$^{3}$.
Linear-response calculations, yielding vibrational and dielectric properties are performed 
using density functional perturbation theory.\cite{LR}

\section{RESULTS}

\subsection{Structural properties}

\begin{table*}[htbp]
\caption{Comparison of the calculated internal atomic parameters with the experimental results from Ref.~\onlinecite{Belokoneva-91}.
The results in italic are theoretical results obtained when fixing the lattice constants at their experimental values.
\label{internal}}
\begin{minipage}[c]{15cm}
\begin{ruledtabular}
\begin{tabular}{lrrrrrr}
&\multicolumn{3}{c}{Theory}&\multicolumn{3}{c}{Experiment}\\
&\multicolumn{1}{c}{$x/a$}&\multicolumn{1}{c}{$y/a$}&\multicolumn{1}{c}{$z/c$}
&\multicolumn{1}{c}{$x/a$}&\multicolumn{1}{c}{$y/a$}&\multicolumn{1}{c}{$z/c$}\\
\hline
La& 0.4127\0\textit{ 0.4120}&-0.0009\0\textit{-0.0009}& 0.3346\0\textit{ 0.3378}& 0.4107 &-0.0007 & 0.333\0\\[0.5mm]
B &-0.1613\0\textit{-0.1145}&-0.0158\0\textit{-0.0143}& 0.3069\0\textit{ 0.3111}&-0.114\0&-0.010\0& 0.313\0\\[0.5mm]
Ge& 0.4191\0\textit{ 0.4171}& 0.0064\0\textit{-0.0041}& 0.8362\0\textit{ 0.8384}& 0.4200 &-0.0033 & 0.834\0\\[0.5mm]
O1& 0.1545\0\textit{ 0.1538}& 0.3495\0\textit{ 0.3474}& 0.0179\0\textit{ 0.0201}& 0.1538 & 0.3438 & 0.0116 \\[0.5mm]
O2& 0.3275\0\textit{ 0.3279}& 0.1415\0\textit{ 0.1414}& 0.9924\0\textit{ 0.9923}& 0.3320 & 0.1450 & 0.9912 \\[0.5mm]
O3& 0.1347\0\textit{ 0.1337}& 0.6101\0\textit{ 0.6082}& 0.3385\0\textit{ 0.3402}& 0.1440 & 0.6125 & 0.3349 \\[0.5mm]
O4& 0.6063\0\textit{ 0.6062}& 0.1570\0\textit{ 0.1522}&-0.3384\0\textit{-0.3334}& 0.6081 & 0.153\0&-0.3341 \\[0.5mm]
O5& 0.0153\0\textit{ 0.0119}& 0.0625\0\textit{ 0.0579}& 0.7731\0\textit{ 0.7807}& 0.0131 & 0.057\0& 0.7807 \\
\end{tabular}
\end{ruledtabular}
\end{minipage}
\end{table*}

The calculated atomic positions (this work) of the non-equivalent atoms (one unit) are shown in 
Table~\ref{internal} along with the existing experimental results. 
The structural parameters of the other atoms in the unit cell 
can still be calculated by invoking symmetry operations. 
The calculated lattice parameters gave values  of $a$=6.957~\AA~and~$c$=6.742~\AA.
These results are in good agreement compared with the reported experimental X-ray single-crystal diffraction values of $a$=7.002~\AA~and~$c$=6.860~\AA\cite{Belokoneva-91} 
and with the high-resolution neutron powder diffraction values of $a$=7.0018~\AA~and~$c$=6.8606~\AA.\cite{Belokoneva-97}
The theoretical results that have been predicted by this investigation underestimate the experimental lattice parameters by less than 2\%, which is a typical LDA error.
Table~\ref{interatomic} shows the inter-atomic distances of anion-oxygen for B and Ge tetrahedra 
as well as La Polyhedra. Both B and Ge tetrahedra 
are connected to O1 and O2. The Ge-O1(O2) are almost the same. A similar behavior is found in the case 
of B-O1(O2). The atoms O3 and O4 represent free vertex of the Ge tetrahedra with distances slightly smaller 
than those of  Ge-O1(O2). The atom O5 is a bridge connecting adjacent B tetrahedra.
Similar to the case of Ge,  the B-O5s distances are found slightly smaller than those of
B-O1(O2). The result makes it easier to distinguish between three different classes of oxygen atoms according to their environment.
The first class consists of O1 and O2, the second consists of O3 and O4,  and the last class is represented by O5 atoms.
The relaxed ionic structure at fixed lattice parameters was also calculated and the results are presented in Tables \ref{internal} and \ref{interatomic}.
In this investigation, it was found that relaxing the ionic position at fixed experimental lattice parameter 
is a better choice than considering the shorter LDA lattice parameters to obtain a vibrational structure in agreement with experimental data.

\begin{table*}[htbp]
\caption{Anion-oxygen interatomic distances (in~{\AA}). Prime 
indices refer to equivalent atoms in the same or adjacent unit cells.
\label{interatomic}}
\begin{minipage}[c]{15cm}
\begin{ruledtabular}
\begin{tabular}{lcccccccccc}
&\multicolumn{2}{c}{La polyhedra}&\multicolumn{1}{c}{Ge tetrahedra}&\multicolumn{1}{c}{B tetrahedra}\\
\hline
Theory @ $V_\mathrm{Th.}$
& O1: 2.719 & O1$^\prime$: 2.606 & O1: 1.730 & O1:\1 1.501\\ 
& O2: 2.692 & O2$^\prime$: 2.699 & O2: 1.732 & O2:\1 1.506\\
& O3: 2.415 & O3$^\prime$: 2.530 & O3: 1.673 & O5:\1 1.448\\
& O4: 2.414 & O4$^\prime$: 2.530 & O4: 1.678 & O5$^\prime$: 1.446\\
& O5: 2.594 \\[2mm]  
Theory @ $V_\mathrm{Expt.}$
& O1: 2.753 & O1$^\prime$: 2.634 & O1: 1.739 & O1:\1 1.507\\
& O2: 2.747 & O2$^\prime$: 2.721 & O2: 1.740 & O2:\1 1.513\\
& O3: 2.435 & O3$^\prime$: 2.574 & O3: 1.682 & O5:\1 1.457\\
& O4: 2.435 & O4$^\prime$: 2.575 & O4: 1.686 & O5$^\prime$: 1.455\\
& O5: 2.635 \\[2mm]
Experiment
& O1: 2.772 & O1$^\prime$: 2.651 & O1: 1.782 & O1:\1 1.488\\
& O2: 2.680 & O2$^\prime$: 2.684 & O2: 1.772 & O2:\1 1.535\\
& O3: 2.433 & O3$^\prime$: 2.566 & O3: 1.625 & O5:\1 1.414\\
& O4: 2.412 & O4$^\prime$: 2.589 & O4: 1.761 & O5$^\prime$: 1.481\\
& O5: 2.672 \\
\end{tabular}
\end{ruledtabular}
\end{minipage}
\end{table*}

\subsection{Electronic properties}
\begin{figure}[htb]
\begin{center}
\includegraphics{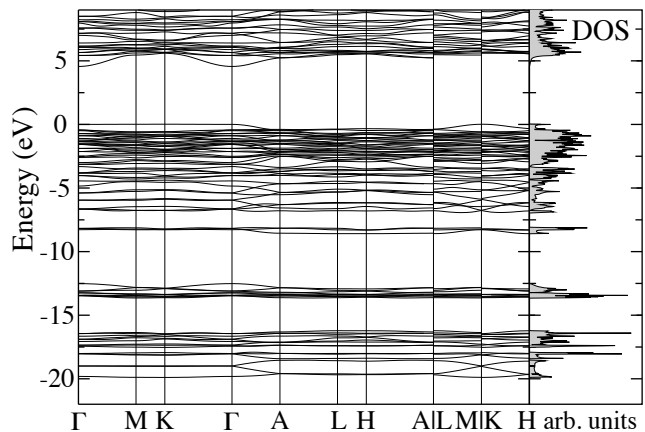}
\end{center}
\caption{Electronic band structure and density of states of LaBGeO$_5$.
The Fermi level has been aligned to the top of the valence
band.\label{bands}}
\end{figure}

The electronic density of states accompanied by the electronic 
band structure along the high symmetry lines~\cite{kpoints} have been computed
as shown in Figure \ref{bands}.
We found that the minimum gap is 
indirect at $\Gamma$-K with a value of 4.54~eV. The minimum direct gap has
almost the same value and exist at $\Gamma$.
However, it is a well-known fact that 
the DFT predictions underestimate the value of the electronic gap. For similar materials,
an underestimation by 2~eV is not uncommon.
A more precise value of the gap can be obtained usually by
employing many-body perturbation theory (the $GW$ approximation), which is however beyond 
the scope of this study. However, although 
the DFT value of the electronic band gap is
known to differ from the experimental one, the characteristics
of the valence band are usually believed faithful.

As seen in Figure \ref{bands}, five groups of bands are identified. The lowest set (not shown, consisting of 3 bands) is located at -28~eV,
characterized by a sharp DOS peak, and attributed to La $s$-states. The second set (consisting of 15 bands), located between
-20 and -16.5 and related to O $s$-bands, is slightly hybridized with Ge, B and La orbitals. The third set (consisting of 9 bands) which is located between
-13.5 and -12.5~eV, corresponds to La $p$-orbitals. The fourth bands (consisting of 3 bands) is located between -8.6 and -8.1~eV and attributed
to B $s$-orbitals. This set is slightly hybridized with O $p$-orbitals. 
The last set, which has the largest dispersion ($\sim$7~eV), is composed of a mix of O $p$-, La $d$-, and Ge $p$- and B $p$-orbitals.
The conduction band edge consists of a mix of La-$s$, La-$d$ and Ge-$s$ with a slight hybridization of O orbitals. 
Analysis of the energy bands indicate a mixed ionic-covalent behavior, as will be confirmed in the following section.

\subsection{Born effective charge}

The Born effective charge tensor $Z^{*}_{\alpha\beta,j}$ is defined as the induced polarization of the solid along the Cartesian direction $\alpha$ by a unit displacement in the direction $\beta$ of the sublattice generated by atom $j$.
Equivalently, it is the force along the direction $\alpha$ on the atom $j$ due to a homogeneous electric field in the direction $\beta$.~\cite{Ghosez} 
The effective charge tensors have been calculated for the eight non-equivalent atoms in the unit cell. 
The results are shown in Table~\ref{born}. The charge tensors for the remaining atoms of the unit cell can be still be obtained 
by invoking the symmetry operations.
The charge neutrality sum rule ($\sum_j Z^{*}_{\alpha\beta,j}$=0) is almost perfectly verified (with a total value lower than $0.05$) 
suggesting that our results are well converged. 

\begin{table*}[htbp]
\caption{Calculated Born effective charge tensors $Z^{*}$ along with their principal values $\lambda$.
The percentages indicate the relative deviations of the principal values from the isotropic values.
\label{born}}
\begin{minipage}[c]{15cm}
\begin{ruledtabular}
\begin{tabular}{lcclcc}
Ion & $Z^{*}$ & $\lambda$ &  Ion & $Z^{*}$ & $\lambda$ \\ 
\hline
La
&  
$ \left( \begin{array}{rrr}
   3.82 & \2 0.06 &  -0.15\\
   0.03 &    4.54 &   0.26\\
  -0.14 &    0.15 &   4.08
  \end{array} \right)$ 
&
$ \begin{array}{rr}
\2 4.61 & 11\% \\
   4.09 & -1\% \\
   3.73 &-10\%  
  \end{array}$
&
Ge
&
$ \left( \begin{array}{rrr}
   3.18   &  -0.01   &  0.07\\
  -0.04   &   2.92   & -0.16\\
   0.02   &  -0.15   &  3.62
  \end{array} \right)$
&
$ \begin{array}{rr}
\2 3.66 & 13\% \\
   3.18 & -2\% \\
   2.89 &-11\% 
  \end{array}$
\\[7mm]
B
&  
$ \left( \begin{array}{rrr}
   2.58    & -0.14   &   0.12\\
  -0.13    &  2.59   &  -0.14\\
   0.17    & -0.12   &   2.48
  \end{array} \right)$
&
$ \begin{array}{rr}
\2 2.83 &\phantom{-}11\% \\
   2.44 &  -4\% \\
   2.38 &  -7\%
  \end{array}$ 
&
O1
&
$ \left( \begin{array}{rrr}
 -2.15  &  0.24   & ~0.29\\
  0.33  & -2.08   & -0.26\\
  0.37  & -0.15   & -1.73
  \end{array} \right)  $
&
$ \begin{array}{rr}
  -2.57 & 30\% \\
  -1.83 & -8\% \\
  -1.55 &-22\%
  \end{array} $ 
\\[7mm]
O2
&
$ \left( \begin{array}{rrr}
  -2.10   & -0.43   & -0.18\\ 
  -0.53   & -2.13   & -0.07\\
  -0.20   &  0.01   & -1.72
  \end{array} \right)  $
&
$ \begin{array}{rr}
  -2.62 & 32\% \\
  -1.78 &-11\% \\
  -1.55 &-22\%
  \end{array} $
&
O3
&
$ \left( \begin{array}{rrr}
  -1.27   & -0.39   & -0.07\\
  -0.37   & -2.60   & -0.14\\
  -0.08   & -0.17   & -2.21
  \end{array} \right)  $
&
$ \begin{array}{rr}
  -2.76 & 36\%\\
  -2.16 &  6\%\\
  -1.17 &-42\%
  \end{array} $
\\[7mm]
O4
&
$ \left( \begin{array}{rrr}
   -1.28   &  0.36   &  0.06\\
    0.37   & -2.63   & -0.10\\
    0.07   & -0.17   & -2.23
    \end{array} \right)  $
&
$ \begin{array}{rr}
  -2.77 & 35\%\\
  -2.19 &  7\%\\
  -1.19 &-42\%
  \end{array} $
&
O5
&
$ \left( \begin{array}{rrr}
   -1.44    & 0.39  &  -0.03\\
    0.35    &-1.95  &  -0.58\\
   -0.02    & 0.54  &  -2.29
   \end{array} \right)  $
&
$ \begin{array}{rr}
  -2.74 & 45\%\\
  -1.78 & -6\%\\
  -1.16 &-39\%
  \end{array} $
\\
\end{tabular}
\end{ruledtabular}
\end{minipage}
\end{table*}

Since the local site symmetry of each ion in 
this structure \textit{can be completely represented by $C_1$} , 
the charge tensors of all ions are anisotropic, with 
non-zero off-diagonal elements. Thus it might be more convenient to 
analyze the charge tensor elements along the ion principal axis (the principal values).

The effective charge of each La-ion, between 3.73 and 4.61, is anomalously large compared to its nominal ionic charges (La: +3). This agrees well with what have been obtained in case of La$_2$O$_3$.~\cite{Hosseini} 
The anomalously large $Z^{*}$ values indicate a strong dynamic
charge transfer along the La–O bond, confirming the above-mentioned a mixed ionic--covalent bond.

For Ge and B ions, the effective charges are noticeably smaller than  
their nominal ionic charges (Ge: +4, B: +3). 
The presence of five non-equivalent oxygen atoms is reflected in differences 
between their Born effective charge tensors. 
Hence, 3 different types of Oxygen atoms are recognized. Where O1 and O2 is each 
connected to Ge and B  tetrahedra simultaneously. O5 is a bridge between two B tetrahedra. 
O3 and O4 are each connected to a single Ge atom.

The anisotropy of the effective charge tensor
is measured by considering the deviation of each principal value 
from the isotropic value. The anisotropy of La-ion charge tensor 
is found to be (11, -1, and -10\%). In case of Ge (13, -2, -11\%), 
and (11, -4, -7\%) for B. The weak anisotropy of La-, Ge-, B-ions can be 
understood in terms of the bonding environment. 
Each La-ion is surrounded by 9 O-ions with close bonding lengths.
 
Similar arguments can be drawn for Ge and B where both are connected by tetrahedral bonds to O-ions. Despite the small differences 
in bond lengths, still the isotropic environment remains a valid scenario.

\subsection{Vibrational and Dielectric properties}

\begin{table*}[htbp]
\caption{Calculated and experimental TO and LO frequencies (in cm$^{-1}$).
The contributions of the phonon modes to the static dielectric
tensor $\Delta\varepsilon$ are indicated.
\label{freq}}
\begin{minipage}[c]{15cm}
\begin{ruledtabular}
\begin{tabular}{lrrrrrrrrrrr}
& \multicolumn{3}{c}{DFPT @ $V_\mathrm{Expt.}$}
& \multicolumn{3}{c}{DFPT @ $V_\mathrm{Th.}$}
& \multicolumn{2}{c}{Model~\cite{Smirnov}}
& \multicolumn{3}{c}{Experiment~\cite{Hruba}}\\
Mode
& TO& LO& $\Delta\varepsilon$
& TO& LO& $\Delta\varepsilon$
& TO& $\Delta\varepsilon$
& TO& LO& $\Delta\varepsilon$ \\
\hline
 $E_1$   &  90&  90&0.05&  89&  89&0.04& 101&1.07&  92&  93&0.27\\     
 $E_2$   & 110& 111&0.13& 113& 114&0.12& 119&1.43& 109& 110&0.29\\     
 $E_3$   & 128& 128&0.15& 133& 133&0.01& 125&1.85& 124& 125&0.34\\     
 $E_4$   & 163& 176&2.86& 162& 180&2.47& 141&0.22& 162& 179&2.99\\     
 $E_5$   & 183& 197&0.56& 196& 199&0.42& 187&0.05& 187& 200&0.48\\     
 $E_6$   & 207& 211&0.11& 214& 214&0.17& 214&0.00& 207& 208&0.05\\     
 $E_7$   & 221& 222&0.04& 234& 235&0.04& 245&0.01& 233& 234&0.08\\
 $E_8$   & 259& 261&0.14& 263& 265&0.14& 289&0.09& 258& 262&0.22\\     
 $E_9$   & 307& 321&0.90& 310& 324&0.81& 319&0.57& 301& 320&1.00\\     
$E_{10}$ & 333& 347&0.29& 345& 363&0.44& 370&0.14& 336& 352&0.29\\     
$E_{11}$ & 377& 391&0.31& 390& 402&0.23& 382&0.21& 384& 396&0.23\\     
$E_{12}$ & 418& 431&0.17& 431& 446&0.19& 456&0.01& 423& 439&0.20\\     
$E_{13}$ & 446& 446&0.00& 458& 460&0.01& 518&0.00&\multicolumn{2}{c}{(454)}& --- \\       
$E_{14}$ & 485& 489&0.06& 500& 503&0.05& 557&0.01& 496& 502&0.08\\     
$E_{15}$ & 609& 611&0.05& 624& 626&0.04& 653&0.00& 616& 621&0.11\\     
$E_{16}$ & 690& 690&0.02& 702& 703&0.04& 719&0.03& 695& 701&0.32\\     
$E_{17}$ & 723& 743&0.32& 737& 756&0.33& 717&0.03& 722& 753&0.33\\     
$E_{18}$ & 790& 824&0.37& 811& 847&0.38& 785&0.04& 784& 811&0.19\\     
$E_{19}$ & 835& 843&0.02& 859& 866&0.02& 809&0.02& 826& 834&0.02\\     
$E_{20}$ & 869& 872&0.02& 892& 895&0.01& 861&0.04& 859& 863&0.02\\     
$E_{21}$ & 912& 923&0.10& 930& 943&0.10& 936&0.04& 918& 928&0.08\\     
$E_{22}$ & 971&1031&0.25& 998&1056&0.24& 968&0.01& 975&1042&0.30\\   
$E_{23}$ &1092&1095&0.00&1120&1124&0.01&1043&0.03&1088&1098&0.01\\ 
\hline
 $A_1$   &  91&  93&0.47&  90&  90&0.00& 105&0.10&  87&  89&0.46\\
 $A_2$   &  97&  97&0.02&  95&  96&0.31& 110&0.16&  95&  96&0.02\\     
 $A_3$   & 118& 122&1.15& 124& 126&0.60& 122&0.45& 117& 119&0.63\\   
 $A_4$   & 148& 149&0.18& 154& 155&0.17& 142&0.07& 144& 145&0.21\\
 $A_5$   & 166& 184&4.46& 186& 196&3.34& 163&1.31& 173& 213&4.28\\
 $A_6$   & 184& 210&0.00& 199& 212&0.07& ---& ---& ---& ---& ---\\ 
 $A_7$   & 210& 217&0.01& 213& 230&0.04& 198&0.01& 215& 222&0.01\\    
 $A_8$   & 280& 280&0.00& 283& 283&0.00& 284&0.00&\multicolumn{2}{c}{(273)}& ---\\       
 $A_9$   & 301& 302&0.04& 306& 307&0.04& 304&0.02&\multicolumn{2}{c}{(301)}& ---\\     
 $A_{10}$& 303& 304&0.01& 319& 319&0.00& 340&0.25& 306& 307&0.05\\   
 $A_{11}$& 322& 327&0.23& 334& 334&0.19& 375&0.00& 324& 329&0.26\\     
 $A_{12}$& 372& 381&0.68& 381& 395&0.82& 401&0.47& 368& 380&0.76\\  
 $A_{13}$& 388& 419&0.34& 403& 433&0.29& 485&0.00& 389& 422&0.34\\     
 $A_{14}$& 492& 501&0.14& 498& 509&0.15& 555&0.00& 503& 510&0.11\\    
 $A_{15}$& 539& 544&0.07& 550& 556&0.07& 599&0.00& 546& 552&0.03\\    
 $A_{16}$& 624& 626&0.03& 637& 640&0.03& 697&0.00& 631& 633&0.03\\   
 $A_{17}$& 732& 744&0.20& 749& 762&0.18& 752&0.04& 733& 745&0.20\\    
 $A_{18}$& 810& 817&0.12& 830& 836&0.08& 792&0.00& 799& 803&0.12\\    
 $A_{19}$& 818& 819&0.00& 840& 840&0.00& 800&0.16& 806& 813&0.04\\    
 $A_{20}$& 850& 854&0.09& 870& 872&0.08& ---& ---& 847& 852&----\\   
 $A_{21}$& 860& 873&0.11& 880& 896&0.13& 848&0.01& 864& 866&0.04\\   
 $A_{22}$& 948& 981&0.43& 986&1005&0.34& 930&0.01& 941& 980&0.51\\   
 $A_{23}$& 990&1043&0.07&1013&1074&0.10&1034&0.07& 992&1050&0.08\\ 
\end{tabular}
\end{ruledtabular}
\end{minipage}
\end{table*}

Group theory analysis, for space group $C_3$, indicates that the irreducible representations of phonon modes at $\Gamma$ are 
\[
\Gamma_\mathrm{vib}=24E\oplus24A ,
\]

where $E$ and $A$ modes are doubly and singly degenerate modes, 
respectively, and generate together a total of 72 modes.
The $E$ modes are marked by collective displacement patterns 
in the $x-y$ plane, while the $A$ modes has 
collective displacements along the $z$ direction. 
Due to absence of the center of inversion,
all modes allow simultaneous IR and Raman activities. 
The contribution of these modes to the frequency dependent dielectric permittivity 
is calculated via

\[
\varepsilon^{0}_{\alpha\beta}(\omega)=\varepsilon^{\infty}_{\alpha\beta}+\sum_{m}\Delta\varepsilon_{\alpha\beta,m}(\omega)
\]

where $\Delta\varepsilon_{\alpha\beta,m}$ is the contribution of the vibrational mode $m$ to the dielectric tensor
along the Cartesian coordinates $\alpha$ and $\beta$ and given as

\[
\Delta\varepsilon_{\alpha\beta,m}(\omega)=\frac{4\pi}{V}\frac{S_{\alpha\beta,m}}{\omega^2_m-\omega^2-i\Omega\omega}
\]

$S_{\alpha\beta,m}$ is the $m^\mathrm{th}$ mode oscillator strength along both $\alpha$ and $\beta$ direction. The damping frequency $\Omega$ 
is taken to be identical for all modes with a value 7~cm$^{-1}$.

The infrared reflectivity is related to the frequency dependent dielectric
permittivity as
\[
R(\omega)=\left|\frac{\sqrt{\varepsilon^{0}_{\alpha\beta}(\omega)}-1}{\sqrt{\varepsilon^{0}_{\alpha\beta}(\omega)}+1}\right|^2
\]

\begin{figure}[htb]
\includegraphics{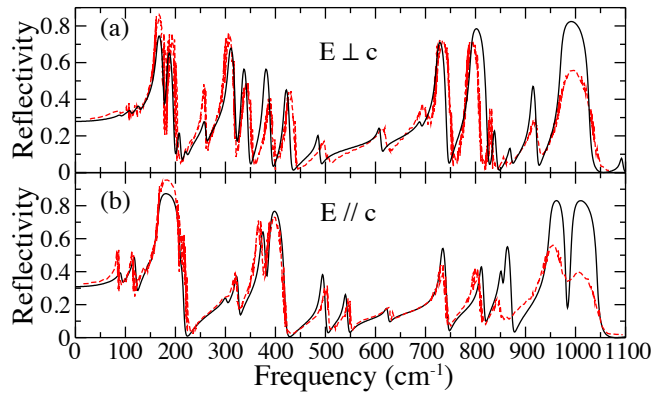}
\caption{ Calculated (black) and measured (red) IR reflectivity parallel and perpendicular to the trigonal axis.
The experimental results are from Ref.~\onlinecite{Hruba}.
\label{Refl}}
\end{figure}

The calculated phonon frequencies
at the center of the Brillouin zone $\Gamma$, computed in this work,
are compared with measured values and presented in Table \ref{freq}.
For comparison reasons, phonon frequencies from theoretical model calculations
are also cited.
For each mode, longitudinal and transverse optical (LO) and (TO) frequencies are being shown. 
Such splitting is normally attributed 
to the long range dipole-dipole
interaction.
Results from this investigation demonstrate good agreement
with existing experimental data with absolute and relative rms values of 6.04~cm$^{-1}$, and 1.81\%, respectively.

Two additional $A$ modes are found at 184~cm$^{-1}$ and 301~cm$^{-1}$, respectively. 
They are marked by their low oscillator strengths and as such introduce a negligible contributions to the 
dielectric permittivity, which explains why they were not detected experimentally. 

Due to the trigonal symmetry of the crystal, the electronic $\varepsilon_\infty$  and static dielectric tensors  $\varepsilon_0$ 
possess two independent components
$\varepsilon_\parallel$ and  $\varepsilon_\perp$, along and perpendicular to the $c$ axis, respectively.

These values of $\varepsilon_\infty$ and $\varepsilon_0$ are presented in Table \ref{dielectric}.
It has been noticed that the major mode contribution to  $\varepsilon_\perp$ comes from $E_4$ mode 
(experimentally observed at 163~cm$^{-1}$), 
with $\Delta\varepsilon$=2.77, that is nearly 40\% of the overall lattice contribution. 
This outcome agrees reasonably well with the experimental findings that $\Delta\varepsilon$ of the equivalent mode is found to be the largest among other modes, with a value around~2.99. Similar agreement has been found in the case of $\varepsilon_\parallel$, the largest ionic contribution coming from the $A_5$ mode 
(experimentally observed at 173~cm$^{-1}$), with
$\Delta\varepsilon$=4.33 in reasonable agreement with the experimental value of 4.28. 
The large contribution from these modes, found experimentally, had not been reproduced by the 
model calculations of Ref. [\onlinecite{Smirnov}]. 
Analyzing the eigendisplacement of the modes with the highest $\Delta\varepsilon$, namely, $E_4$ and $A_5$,
we found that these modes are characterized by rigid unit translations of Ge and B tetrahedra opposite to La ions.

\begin{table}[htbp]
\caption{The dielectric permittivity tensor components along and perpendicular to the trigonal axis.
The sums of ionic contributions to static dielectric tensor $ \Delta\varepsilon_{\mathrm{tot}} $ are also indicated. 
\label{dielectric}}
\begin{minipage}[c]{4cm}
\begin{ruledtabular}
\begin{tabular}{lrr}
& \multicolumn{1}{c}{$\varepsilon_\parallel$} & \multicolumn{1}{c}{$\varepsilon_\perp$}\\[1
mm]
\hline
$\varepsilon_\infty$&3.55&  3.71\\
$\Delta\varepsilon_{\mathrm{tot}}$ &6.79&  8.57\\
\hline
    $\varepsilon_0$ &10.34&12.28\\
\end{tabular}
\end{ruledtabular}
\end{minipage}
\end{table}

In general we found that the low-frequency modes have significant contributions to the dielectric tensor due
to the large displacement of the La ions (for which the Born effective charge is anomalously large, as mentioned in the previous section). This is in contrast with high-frequency modes which have less contribution 
to the dielectric tensor, due to negligible displacements of the La ions.

\begin{figure}[htb]
\includegraphics{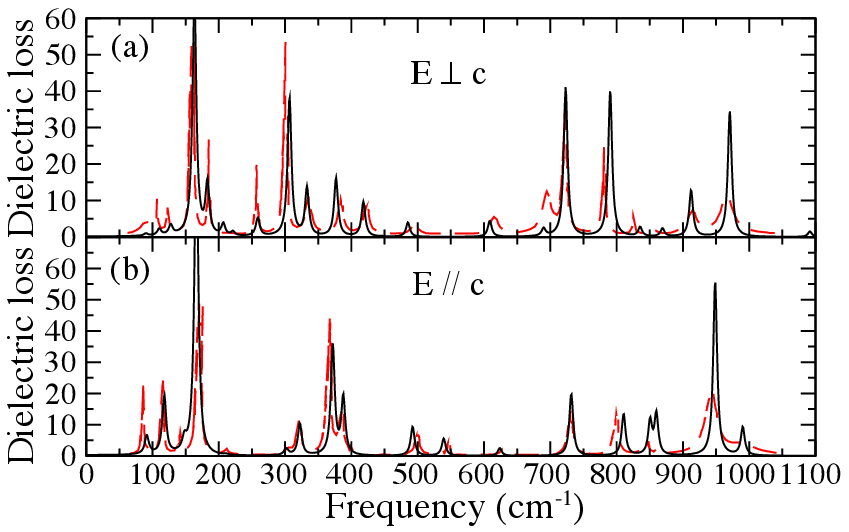}
\caption{Calculated (black) and measured (red) dielectric loss parallel and perpendicular to the trigonal axis.
The experimental results are from Ref.~\onlinecite{Hruba}.
\label{loss}}
\end{figure}

The calculated IR reflectivity and dielectric loss 
are presented in Figs.~\ref{Refl} and~\ref{loss} respectively, along with the correspondent measured data.
The overall agreement between both calculated and experimental results is rather good as all
the major features of the experimental infrared spectra are already obtained.

For comparison reasons, it is noted that $\varepsilon_\infty$ can be estimated from the refraction index $n$, which has been reported
experimentally to be about 1.88,~\cite{Kaminskii} $\varepsilon_\infty$=$n^2$=3.53. The calculated values of $\varepsilon_\infty$ are 3.55 
and 3.71 for the perpendicular and parallel components, respectively. The anisotropy is weak, although non-negligible.
They are in excellent agreement with experimental measurements.

In fact, the  reported value of $\varepsilon_0$=11
falls nicely between the presently calculated values of 10.34 and 12.28 
for the perpendicular and parallel components, respectively. The ionic contribution to $\varepsilon_0$ is much more anisotropic
than the electronic contribution.

\section{Conclusion}
A comprehensive study of the structural, electronic, 
vibrational and dielectric properties of the LaBGeO$_5$ 
compound with the stillwellite structure has been 
presented in this investigation using ab initio density functional theory. 
The structural parameters of this compound were found to agree well with 
the available experimental data with a negligible deviation of less than 0.2\%.
The DFT electronic structure, which had not yet been computed, has been presented. 
The calculated vibrational and dielectric properties of this representative compound of the larger rare-earth borogermanate family  of the stillwellite structure are found  to fit neatly and therefore 
consolidate the existing experimental published results.
In particular, we improve the agreement significantly with respect to the previous model calculation.

\begin{acknowledgments}
The authors RS, BH and JK acknowledge the generous support of the Scientific Research Support Fund (SRF) from the Ministry of Higher Education and Research in Jordan. 
R.S. acknowledge support from the U.C.L. for a stay in Louvain-la-Neuve (Belgium). 
Computational ressources have been provided by the supercomputing facilities of the Universit\'e catholique de Louvain (CISM/UCL)
as well as from the Consortium des equipements de Calcul Intensif en F\'ed\'eration Wallonie Bruxelles (CECI) that is funded 
by the Fonds de la Recherche Scientifique de Belgique (FRS-FNRS).
\end{acknowledgments}

\bibliography{basename of .bib file}

\end{document}